\documentstyle[aps,preprint]{revtex}
\begin{document}
\preprint{SNUTP-97-036,~gr-qc/9703064}
\title{Coherence and Emergence of Classical Spacetime}

\author{Sang Pyo Kim\footnote{E-Mail: sangkim@knusun1.kunsan.ac.kr}}

\address{Department of Physics\\
Kunsan National University\\
Kunsan 573-701, Korea}

\author{Jeong-Young Ji\footnote{E-Mail: jyji@phyb.snu.ac.kr},
Hak-Soo Shin
and Kwang-Sup Soh\footnote{E-Mail: kssoh@phya.snu.ac.kr}}

\address{Department of Physics Education\\
Seoul National University\\
Seoul 151-742, Korea}

\maketitle

\begin{abstract}
Using the coherent-state representation we show that
the classical Einstein equation for the FRW cosmological model
with a general minimal scalar field can be derived from the semiclassical
quantum Einstein equation.
\end{abstract}
\vspace{1.5cm}

\newpage

Canonical quantum gravity has been initiated by DeWitt in the seminal paper
\cite{dewitt}. As a methodology to understand the quantum aspects of cosmology,
quantum cosmology has been intensively studied, and in particular, as a great
conceptual advancement, the boundary conditions have been incorporated
for the Universe by Hartle-Hawking \cite{hawking} and Vilenkin \cite{vilenkin}.
Semiclassical quantum gravity has also been elaborated as a methodology
to include some part of quantum effects into classical gravity \cite{kiefer}.
In order to consider the different mass scales between gravity and matter
fields and to apply quantum cosmology to the early Universe,
one should have the reduction scheme from canonical
quantum gravity, $\hat{G}_{\mu \nu} = 8 \pi \hat{T}_{\mu \nu}$,
to semiclassical quantum gravity, $G_{\mu \nu} = 8 \pi
\langle \hat{T}_{\mu \nu} \rangle$, and down to classical gravity,
$G_{\mu \nu} = 8 \pi T_{\mu \nu}$.

In this {\it Brief Report}, we complete the reduction from semiclassical
quantum gravity to classical gravity for a quantum FRW cosmological
model with a general minimal scalar field. We find that
the coherent-state representation of the semiclassical quantum Einstein
equation leads to the classical Einstein equation with a quantum correction.
In previous papers \cite{kim1,kim2}, we showed that in the case of
a massive scalar field an exact quantum state of time-dependent
Schr\"{o}dinger equation gives rise to the mean energy density
which has the same form as the classical one except
that the field intensity is replaced by the absolute value,
and that a coherent state exactly gives rise to the classical density
plus an additional one from vacuum fluctuation.
We extend the result of the massive scalar-field model to the general
scalar-field model.

As a quantum cosmological model, we consider the FRW Universe whose
Wheeler-DeWitt equation is given by
\begin{equation}
\left[ \frac{2 \pi \hbar^2}{3 m_P a} \frac{\partial^2}{\partial a^2}
- \frac{3 m_P}{8 \pi} ka - \frac{\hbar^2}{2 a^3} \frac{\partial^2}{\partial
\phi^2}
+ a^3 V(\phi) \right] \Psi (a, \phi) = 0.
\end{equation}
Here, $k$ takes $1, 0$, and $ -1$ for a closed, spatially flat,
and open universe, respectively. The unit system is $c = 1$
and $G = \frac{1}{m_P}$.
The corresponding classical Einstein equation is
\begin{equation}
\Bigl(\frac{\dot{a}}{a} \Bigr)^2 + \frac{k}{a^2} =
\frac{8 \pi}{3 m_P} \Bigl( \frac{\dot{\phi}^2}{2} + V(\phi) \Bigr),
\end{equation}
and the classical field equation is
\begin{equation}
\ddot{\phi} + 3 \frac{\dot{a}}{a} \dot{\phi} + \frac{d V(\phi)}{d \phi} = 0.
\end{equation}

Following the reduction scheme, one obtains the semiclassical
quantum gravity from the Wheeler-DeWitt equation:
\begin{equation}
\Bigl(\frac{\dot{a}}{a} \Bigr)^2 + \frac{k}{a^2} =
\frac{8 \pi}{3 m_P a^3} \langle \hat{H} \rangle,
\end{equation}
and
\begin{equation}
i \hbar \frac{\partial}{\partial t} \Phi (\phi, t) = \hat{H} \Phi (\phi,t)
\end{equation}
where
\begin{equation}
\hat{H} = \frac{1}{2 a^3} \hat{\pi}^2 + a^3 V(\hat{\phi}).
\end{equation}

We now represent the semiclassical Einstein equation in the coherent state.
For the case of the massive scalar-field model, the coherent-state
representation was given explicitly in terms of classical solutions
\cite{kim2}.
For the case of a general scalar-field model, we use the coherent states
constructed by Rajagopal and Marshall \cite{rajagopal}. We follow their main
idea but redefine some of variables to be suitable for the application
to quantum field theory in FRW cosmology.
We construct the Fock space by introducing  the creation and
annihilation operators
\begin{eqnarray}
\hat{A}^{\dagger} (t) = -i[ u (t) \hat{\pi} - a^3 \dot{u} (t) \hat{\phi}],
\nonumber\\
\hat{A} (t) = i[ u^* (t) \hat{\pi} - a^3 \dot{u}^* (t) \hat{\phi}].
\end{eqnarray}
As in Ref. \cite{kim3}, if we require that $\hat{A}^{\dagger}$ and
$\hat{A}$ be the invariant operators
\begin{equation}
i \hbar \frac{\partial}{\partial t}
\{ \begin{array}{clcr} \hat{A}^{\dagger} \\  \hat{A} \end{array} \} +
[ \{ \begin{array}{clcr} \hat{A}^{\dagger} \\ \hat{A} \end{array} \} ,
\hat{H}] = 0,
\end{equation}
then $u$ satisfies the equation
\begin{equation}
(a^3 \dot{u} ) \dot{} \hat{\phi} + a^3 u V^\prime (\hat{\phi}) = 0 .
\label{u:ph}
\end{equation}
{}From the usual commutation relation it follows that
\begin{equation}
\hbar a^3 \Bigl( u  \dot{u}^*  - u^*  \dot{u}  \Bigl) = i.
\end{equation}
The one parameter-dependent vacuum is defined to be annihilated
by the annihilation operator
\begin{equation}
\hat{A}(t) \vert 0, t \rangle = 0.
\label{vac}
\end{equation}
The momentum and position operators are given by
\begin{eqnarray}
\hat{\phi} = \hbar (
u  \hat{A} + u^* \hat{A}^{\dagger} ),
\nonumber\\
\hat{\pi} = \hbar a^3 (
\dot{u} \hat{A} + \dot{u}^* \hat{A}^{\dagger}).
\end{eqnarray}
A coherent state is an eigenstate of $ \hat{A} (t)$:
\begin{equation}
\hat{A} (t) \vert \alpha, t \rangle = \alpha \vert \alpha, t \rangle.
\end{equation}
It is a unitary transformation of the vacuum by a displacement operator
\begin{equation}
\vert \alpha, t \rangle = \exp \Bigl( \alpha \hat{A}^{\dagger} - \alpha^*
\hat{A}  \Bigr) \vert 0, t \rangle.
\end{equation}
The position and momentum operators in the coherent-state representation
yield the classical field and momentum
\begin{eqnarray}
\phi_c \equiv \langle \alpha, t \vert \hat{\phi}
\vert \alpha, t \rangle =
\hbar ( u  \alpha + u^* \alpha^* ),
\nonumber\\
\pi_c \equiv \langle \alpha, t \vert \hat{\pi}
\vert \alpha, t \rangle =
\hbar a^3 (
\dot{u} \alpha + \dot{u}^* \alpha^* ).
\end{eqnarray}
Thus, in the coherent-state representation, the semiclassical Einstein
equation leads to
\begin{eqnarray}
\Bigl(\frac{\dot{a}}{a} \Bigr)^2 + \frac{k}{a^2} &=&
\frac{8 \pi}{3 m_P a^3} \langle \alpha, t \vert \hat{H}
\vert \alpha, t \rangle \nonumber\\
&=& \frac{8 \pi}{3 m_P a^3} \Bigl( \frac{\pi_c^2}{2 a^3}
+ a^3 V(\phi_c)  + H_{q} \Bigr)
\label{sem ein}
\end{eqnarray}
where
\begin{equation}
H_{q} =  \frac{\hbar^2}{2} a^3 \dot{u}^* \dot{u} +
a^3 \Bigl[ \exp \Bigl( \frac{\hbar^2}{2} u^* u
\frac{\partial^2}{\partial \phi_c^2} \Bigr)
- 1 \Bigr]  V(\phi_c)
\label{quan cor}
\end{equation}
is a quantum correction to the classical energy density.
By identifying $\pi_{u} = a^3 \dot{u}^*$ and $\pi_{u^*} = a^3 \dot{u}$,
one may obtain the Hamilton equations for $H_{q}$
\begin{equation}
\bigl( a^3 \dot{u} \bigr)\dot{} +
\frac{2}{\hbar^2} a^3 \frac{\partial}{\partial u^*}
\Bigl[ \exp \Bigl( \frac{\hbar^2}{2} u^* u
\frac{\partial^2}{\partial \phi_c^2} \Bigr)
- 1 \Bigr] V(\phi_c)  = 0.
\label{mean eq}
\end{equation}
The equation (\ref{mean eq}) for $u$ is identical to the equation obtained
by differentiating Eq. (\ref{u:ph}) with respect to $\hat{\phi}$  and
taking the expectation value with the coherent state $\vert \alpha, t
\rangle$.
The coincidence of the invariant equation and the
Hamilton equations in the same Fock-space representation
follows from the physical principle of the extremization
of action \cite{balian} for any time-dependent system.
It should be noted that the quantum corrections
are of the order of $\hbar^2$ or higher.

Finally, we compare the semiclassical Einstein equation in the
coherent state representation (\ref{sem ein}) with that
in a different quantum state of matter field.
As one of the non-coherent states,
we consider the vacuum state (\ref{vac}) of the Fock space.
The vacuum  state can be regarded as a specific coherent state with
$\phi_c = \pi_c = 0$. Thus, the non-zero contributions
come from quantum fluctuations around $\phi =0$:
\begin{eqnarray}
\Bigl(\frac{\dot{a}}{a} \Bigr)^2 + \frac{k}{a^2} &=&
\frac{8 \pi}{3 m_P a^3} \langle 0, t \vert \hat{H}
\vert 0, t \rangle \nonumber\\
&=& \frac{8 \pi}{3 m_P a^3} \Biggl[
\frac{\hbar^2}{2} a^3 \dot{u}^* \dot{u} +
a^3 \Bigl[ \exp \Bigl( \frac{\hbar^2}{2} u^* u
\frac{\partial^2}{\partial \phi_c^2} \Bigr)
- 1 \Bigr]  V(\phi_c = 0) \Biggr].
\label{vac sem}
\end{eqnarray}

One noticeable difference between the two semiclassical Einstein equations
is that in Eq. (\ref{sem ein}) the coherent field $\phi_c$ can oscillate,
whereas in Eq. (\ref{vac sem}) there is no oscillating behavior
of the vacuum expectation value, since it depends only
the magnitude of field $u$. In cosmological models, $\phi_c$ acts as an
inflaton, whose coherent oscillation
may play an important role in the preheating mechanism.

In summary, we showed that for the FRW cosmological model with
a general minimal scalar field, the semiclassical Einstein equation
(\ref{sem ein}) can reduce to the classical Einstein equation
whose energy density is the sum of classical one and the quantum
correction (\ref{quan cor}). Therefore, we suggest that
coherence may be one of the necessary mechanisms for the emergence
of classical spacetime from the semiclassical quantum gravity. Though
not shown explicitly, the coherence can be understood partially
from the dissipation of quantum fields as the Universe expands.

\section*{acknowledgments}
JYJ, SPK, and HSS were supported by the Non-Directed Research Fund,
Korea Research Foundation, 1996, and KSS was supported
by the Center for Theoretical Physics, the Seoul National University.


\begin{references}
\bibitem{dewitt} B. S. DeWitt, Phys. Rev. {\bf 160}, 1113 (1967).
\bibitem{hawking} J. B. Hartle and S. W. Hawking, Phys. Rev. D {\bf 28},
2960 (1983); S. W. Hawking, Nucl. Phys. {\bf B239}, 257 (1984).
\bibitem{vilenkin} A. Vilenkin, Phys. Rev. D {\bf 27}, 2848 (1983);
{\bf 30}, 509 (1984); Nucl. Phys. {\bf B252}, 141 (1985).
\bibitem{kiefer} For review and references, see
C. Kiefer, in {it Canonical Gravity: From Classical to
Quantum}, edited by J. Ehlers and H. Friedrich (Springer, Berlin, 1994).
\bibitem{kim1} S. P. Kim, J. Korean Phys. Soc. {\bf 28}, S512 (1995);
Phys. Rev. D {\bf 52}, 3382 (1995); Phys. Lett. A {\bf 205}, 359 (1995).
\bibitem{kim2} S. P. Kim, Class. Quantum Grav. {\bf 13}, 1377 (1996);
J. Korean Phys. Soc. {\bf 30}, 349 (1997); Phys. Rev. D 55, (in press).
\bibitem{rajagopal} A. K. Rajagopal and J. T. Marshall,
Phys. Rev. A {\bf 26}, 2977 (1982).
\bibitem{kim3} S. P. Kim, "Mean-Field Approach to Quantum Duffing Oscillator",
physics/9702018.
\bibitem{balian} R. Balian and M. V\'{e}n\'{e}roni, Ann. Phys.
{\bf 164}, 334 (1985).
\end{references}
\end{document}